\documentclass[twocolumn]{revtex4}
\newcommand{\be}{\begin{equation}}
\newcommand{\ee}{\end{equation}}
\usepackage{graphicx}
\usepackage{amsmath}
\begin{document}

% \draft
\preprint{
\vbox{
\hbox{ADP-05-14/T624}
\hbox{DESY 05-183}
}}

\pacs{11.15.Ha, 12.38.Gc, 14.40.Cs}
\title{$1^{-+}$ exotic meson at light quark masses}

\author{J.N.~Hedditch}
\author{W.~Kamleh}
\author{B.G.~Lasscock}
\author{D.B.~Leinweber}
\author{A.G.~Williams}

\affiliation{Department of Physics and Mathematical Physics and\\
	Special Research Centre for the
	Subatomic Structure of Matter,				\\
	University of Adelaide, 5005, Australia}

\author{J.M.~Zanotti}
\affiliation{John von
    Neumann-Institut f\"ur Computing NIC, \\
    Deutches Elektronen-Synchrotron DESY, \\ 
    D-15738 Zeuthen, Germany}

\begin{abstract}
The mass of the $1^{-+}$ exotic meson, created with hybrid interpolating fields,
is explored in numerical simulations of quenched QCD on large ( $20^3 \times 40$ )
lattices to obtain good control of statistical and finite volume errors. Using the
Fat-Link Irrelevant Clover (FLIC) fermion action, the properties of the $1^{-+}$
are investigated at light quark masses approaching 25 MeV ($m_\pi / m_\rho \simeq 1/3$). 
Under the standard assumption that the coupling to the quenched $a_1 \eta^{\prime}$ 
channel comes with a negative metric, our results indicate that the $1^{-+}$ exotic
exhibits significant curvature close to the chiral limit, suggesting previous linear
extrapolations have overestimated the mass of the $1^{-+}$.  We find for the first
time in lattice studies a $1^{-+}$ mass consistent with the $\pi_1 (1600)$ candidate.
We also find a strangeness $\pm 1$ $J^{P} = 1^{-}$ state with a mass close to $2$ GeV.
\end{abstract}
\maketitle

%%%%%%%%%%%%%%%%%%%%%%%%%%%%%%%%%%%%%%%%%%%%%%%%%%%%%%%%%%%%%%%%%%%%%%%%%%
\section{Introduction}

The masses of the so-called `exotic' mesons are attracting
considerable attention from the experimental
community \cite{Alde:1988bv,Tuan:1988ik,Prokoshkin:1994nr,%
Godfrey:1998pd,Chung:1999we,Lu:2004yn,Dzierba:2005sr,HallD} as a
vehicle for the elucidation of the relatively unexplored role of
gluons in QCD.  The Particle Data Group \cite{Eidelman:2004wy} reports two
candidates for the $1^{-+}$ exotic, the $\pi_1 (1400)$ at $1.376(17)\ 
{\rm GeV}$, and the $\pi_1 (1600)$ at $1.596^{+25}_{-14}\ {\rm GeV}$.
The experimental status of these states is an issue that continues to
attract attention \cite{Lu:2004yn,Dzierba:2005sr}.

Lattice QCD provides a first principles approach to nonperturbative
calculations of QCD, indispensable in determining the hadron mass
spectrum.  In the early work of Ref.~\cite{Lacock:1996vy}, the UKQCD
Collaboration made use of gauge-invariant non-local operators to
explore $P$ and $D$-wave mesons, as well as exotics. They used a
tadpole-improved clover action, with 375 configurations on a
$16^3\times48$ lattice.  Their calculation was performed at a single
quark mass corresponding to approximately that of the strange quark
and reported a $1^{-+}$ exotic mass of $1.9(4)$ GeV.

In 1997, the MILC Collaboration used local operators formed by
combining the gluon field strength tensor and standard quark bilinears
\cite{Bernard:1997ib}, the same approach we take in this
paper.  The highly anisotropic lattices employed allowed many time
slices to be used in determining the mass of the exotic.  
$20^3\times48$ and $32^3\times64$ lattices with multiple fermion
sources per lattice were considered.  Using the Wilson action, they
suggested a $1^{-+}$ mass of $1.97(9)$ GeV from a linear
extrapolation in $1/\kappa$ from  $m_\pi^2 \ge 0.64$ GeV.

The SESAM collaboration \cite{Lacock:1998be} analyzed dynamical
fermion configurations at four quark masses corresponding to
$m_\pi^{2} \ge 0.455(9)\ {\rm GeV}^2$.  Their linear extrapolation
resulted in $1.9(2)$ GeV for the mass of the $1^{-+}$ exotic where the
scale was determined via the Sommer parameter~\cite{Sommer:1993ce} $r_0$ via Ref.~\cite{Bali:2000vr}.

Further work using the Clover action, but this time with local
interpolators was performed by Mei \emph{et al.} in Ref.~\cite{Mei:2002ip}.
Very heavy quark masses were used to get good control of statistical
errors.  Their linear extrapolation from $m_\pi^2 \ge 1.05\ {\rm
GeV}^2$ suggested a mass of $2.01(7)$ GeV.

In 2002 the MILC Collaboration published new work
\cite{Bernard:2002rz} using dynamical improved Kogut-Susskind fermions
on $20^3\times48$ and $32^3\times64$ lattices.  Linearly extrapolating
from $m_{\pi}^2 \ge 0.488\ {\rm GeV}^2$, Bernard {\emph et al.}
quote two sets of results for the $1^{-+}$ mass.  Using $r_1 = 0.34$
fm and $\sqrt{\sigma} = $ 440 MeV they report a $1^{-+}$ mass of $1.85(7)$
and $2.03(7)$ GeV respectively.

Michael~\cite{Michael:2003xg} provides a good summary of work to 2003,
concluding that the light-quark exotic is predicted by lattice studies
to have a mass of $1.9(2)$ GeV.

In order to minimize the need for extrapolation one requires access to
quark masses near the chiral regime on large physical volumes.  Our
study considers a physical volume of $(2.6\ \rm fm)^3$, and the ${\mathcal
O}(a)$-improved FLIC fermion action
\cite{Zanotti:2001yb,Zanotti:2002ax,Zanotti:2004dr} whose improved chiral properties
\cite{Boinepalli:2004fz} permit the use of very light quark masses
which are key to our results.

%%%%%%%%%%%%%%%%%%%%%%%%%%%%%%%%%%%%%%%%%%%%%%%%%%%%%%%%%%%%%%%%%%%%%%%%%%
\section{Vector Mesons on the Lattice}

\begin{table*}[hbt] 
\caption{$J^{PC}$ quantum numbers and their associated meson
  interpolating fields. }
\begin{ruledtabular}
\begin{tabular}{cccccccc}
$0^{++}$ & $0^{+-}$ & $0^{-+}$ & $0^{--}$ & $1^{++}$ & $1^{+-}$ & $1^{-+}$ & $1^{--}$\\
\hline
%first line
$\bar{q}^a q^a$ 
& $ i \bar{q}^a \gamma_5 \gamma_j B^{ab}_j q^b$ 
&  $\bar{q}^a \gamma_5 q^a$ 
& $ \bar{q}^a \gamma_5 \gamma_j E^{ab}_j q^b$
& $  \bar{q}^a \gamma_5 \gamma_j q^a$
& $  \bar{q}^a \gamma_5 \gamma_4 \gamma_j q^a$
& $\bar{q}^a \gamma_4 E^{ab}_j q^b$
& $  i \bar{q}^a \gamma_5 B^{ab}_j q^b $\\
%second line
$  \bar{q}^a \gamma_j E^{ab}_j q^b$ 
& $\bar{q}^a \gamma_4 q^a$ 
& $\bar{q}^a \gamma_5 \gamma_4 q^a$
& 
& $ i \bar{q}^a \gamma_4 B^{ab}_j q^b$
& $\bar{q}^a \gamma_5 \gamma_4 E^{ab}_j q^b$
& $i \epsilon_{jkl} \bar{q}^a \gamma_k B^{ab}_l q^b$
& $ \bar{q}^a \gamma_4 \gamma_j q^a$\\
%third line
$ i \bar{q}^a \gamma_j \gamma_4 \gamma_5 B^{ab}_j q^b$ 
& 
& $ i \bar{q}^a \gamma_j B^{ab}_j q^b$
& 
& $  \epsilon_{jkl} \bar{q}^a \gamma_k E^{ab}_l q^b$
& $\bar{q}^a \gamma_5 E^{ab}_j q^b $
& $i \epsilon_{jkl} \bar{q}^a \gamma_4 \gamma_k B^{ab}_l q^b$
& $\bar{q}^a E^{ab}_j q^b$\\
%fourth line
$  \bar{q}^a \gamma_j \gamma_4 E^{ab}_j q^b$ 
& 
& $ i \bar{q}^a \gamma_4 \gamma_j B^{ab}_j q^b$
& 
&  $  \epsilon_{jkl} \bar{q}^a \gamma_k \gamma_4 E^{ab}_l q^b$
& $i \bar{q}^a B^{ab}_j q^b $
& $  \epsilon_{jkl} \bar{q}^a \gamma_5 \gamma_4 \gamma_k E^{ab}_l q^b$
& $  \bar{q}^a \gamma_j q^a$\\
%fifth line
{} &{} &{} &{} &{} &{} &{} &{} $i \bar{q}^a \gamma_4 \gamma_5 B^{ab}_j q^b$  \\
{} &{} &{} &{}\\[-1.5ex]
\end{tabular}\label{interpolators}
\end{ruledtabular}
\end{table*}

Consider the momentum-space meson two-point function
for $t > 0$,
\begin{equation}
{G}^{ij}_{\mu\nu}(t,\vec{p}) = \sum_{\vec{x}} e^{-i\vec{p} \cdot \vec{x}}
\langle\Omega|
  \chi^i_\mu (t,\vec{x})\, {\chi^j_\nu}^{\dagger} (0,\vec 0)
|\Omega\rangle\ 
\end{equation} where $i,j$ label the different interpolating fields
and $\mu , \nu$ label the Lorentz indices.
At the hadronic level,
\begin{eqnarray*}
{G}^{ij}_{\mu\nu}(t,\vec{p})\!\! &=&\!\! \sum_{\vec{x}} e^{-i\vec{p} \cdot \vec{x}}
\sum_{E, \vec{p}^{ \prime}, s }
\langle\Omega| \chi^i_{\mu}(t,\vec{x})|E, \vec{p}\,', s \rangle \nonumber \\
\!\!&\times&\!\!
\langle E, \vec{p}\,', s | {\chi^{j}_{\nu}}^{\dagger}(0,\vec 0) 
|\Omega\rangle\ 
\end{eqnarray*}
where the $|E, \vec{p}\,', s \rangle$ are a complete set of hadronic states,
of energy $E$, momentum $\vec{p}\,'$, and spin $s$,
\begin{equation}
\sum_{ E, \vec{p}^{\prime}, s}
 |E, \vec{p}\,', s\rangle\langle E,  \vec{p}\,', s|=I\ .
\end{equation}
We denote the vacuum couplings as follows:
\begin{eqnarray*}
\langle\Omega|\, \chi^i_\mu \,| E, \vec{p}\,', s \rangle  &=& \lambda^i_E \, \epsilon_{\mu} ({p}\,', s)\,  \\
\langle E, \vec{p}\,', s |\, {\chi^{j}_{\nu}}^{\dagger}\, | \Omega\rangle  &=&  {\lambda^{j}_E}^{\star} \, \epsilon^{\star}_{\nu} ({p}\,', s) \ ,
\end{eqnarray*}
where the four-vector $p\,' = (E, \vec{p}\,')$ is introduced.

We can translate the sink operator from $x$ to $0$ to write this as
\begin{eqnarray}
&& \sum_{\vec{x},E,\vec{p}\,',s} e^{-i\vec{p} \cdot \vec{x}} 
\langle\Omega|\,
  \chi^i_\mu(0)\,
  e^{i\hat{\vec{P}} \cdot \vec{x} -\hat{H}t} 
\,| E, \vec{p}\,', s \rangle \nonumber \\
&& \, \times \, \langle  E, \vec{p}\,', s |\, {\chi^j_\nu}^{\dagger}(0)
\,| \Omega\rangle                            \nonumber \\
&=& \sum_{E, s} e^{-{E t}}
\langle\Omega| \chi^i_\mu | E, \vec{p}, s \rangle
 \langle E, \vec{p}, s | {\chi^{j}_\mu}^{\dagger}|\Omega\rangle\ \nonumber \\
&=& \sum_{E, s} e^{-{E t}}
\lambda^i \epsilon_{\mu} ({p}, s) 
{\lambda^j}^{\star} \epsilon^{\star}_{\nu} ({p}, s)\ .
\end{eqnarray}

We shall label the states
which have the $\chi$ interpolating field quantum numbers
as $|{\alpha}\rangle$ for
$\alpha=1,2,\cdots,N$, thus replacing $\sum_{E}$ with $\sum_\alpha$.  In general the number of states,
$N$, in this tower of excited states may be very large, but we
will only ever need to consider a finite set of the lowest energy
states here, as higher states will be exponentially suppressed as
we evolve to large Euclidean time.
Finally, the transversality condition:
\begin{equation}
\sum_s \epsilon_{\mu} (p, s)\, \epsilon^{\star}_{\nu} (p, s) = - \left(g_{\mu\nu} - \frac{p_\mu p_\nu}{m^2} \right) 
\end{equation} 
implies that for $\vec{p} = 0$, we have
\begin{eqnarray}
\label{eqn:Gijequation}
&&{G}^{ij}_{00}(t,\vec{0}) = 0 \nonumber \\
&&{G}^{ij}_{kl}(t,\vec{0}) = \sum_\alpha \delta_{kl}\, \lambda^i_\alpha \, {\lambda^j}^{\star}_\alpha \, e^{-{m_\alpha t}}\ .
\end{eqnarray}
Since $G^{ij}_{11},G^{ij}_{22}$,and $G^{ij}_{33}$ are all estimates for the same quantity
we add them together to reduce variance, forming the sum
\begin{displaymath}
G^{ij} = G^{ij}_{11} + G^{ij}_{22} + G^{ij}_{33}\ .
\end{displaymath}

Evolving to large Euclidean time will suppress higher mass states exponentially
with respect to the lowest-lying state, leading to the following definition of
the effective mass
\begin{equation}
M^{ij}_{\rm eff}(t) = \ln\left( \frac{G^{ij}(t,\vec{0})}{G^{ij}(t+1,\vec{0})} \right) \ .
\end{equation}
The presence of a plateau in $M_{\rm eff}$ as a function of time, then, signals
that only the ground state signal remains.

%%%%%%%%%%%%%%%%%%%%%%%%%%%%%%%%%%%%%%%%%%%%%%%%%%%%%%%%%%%%%%%%%%%%%%%%%%
\section{Lattice Simulations}

\subsection{Interpolating Fields}

The formulation of effective interpolating fields for the creation and
annihilation of exotic meson states continues to be an active area of
research.  For example, one can generalize the structure of the
interpolating fields to include non-local components where link paths
are incorporated to maintain gauge invariance and carry the nontrivial
quantum numbers of the gluon fields \cite{Lacock:1996vy,Lacock:1998be}.
In this case, numerous quark propagators are required for each gauge
field configuration rendering the approach computationally expensive.
The use of non-local interpolating fields affords greater freedom in
creation of operators and facilitates access to exited states through
variational techniques, but does not lead to an increase in signal
for the ground state $1^{-+}$ exotic commensurate with the increased
computational cost of this approach.  Exotic quantum numbers may also
be obtained from four-quark ($q\bar{q}q\bar{q}$) operators, but in
practice, these tend to exhibit larger statistical fluctuations
\cite{Bernard:1997ib,Lacock:1996vy,Lacock:1998be}.

We consider the local interpolating fields summarized in
Table~\ref{interpolators}.  Gauge-invariant Gaussian smearing
\cite{Gusken:1989qx,Zanotti:2003fx} is applied at the fermion source
($t=8$), and local sinks are used to maintain strong signal in the
two-point correlation functions.  Chromo-electric and -magnetic fields
are created from 3-D APE-smeared links \cite{ape} at both the source
and sink using the highly-improved ${\mathcal O}(a^4)$-improved
lattice field strength tensor \cite{Bilson-Thompson:2002jk} described
in greater detail below.  

While all four $1^{-+}$ interpolating fields of
Table~\ref{interpolators} have been used to create the $1^{-+}$ state
on the lattice, in this work we will focus on the results for the two
$1^{-+}$ interpolating fields coupling large spinor components to
large spinor components: 
\begin{eqnarray}
\chi_2 &=& i \epsilon_{jkl} \bar{q}^a \gamma_k B^{ab}_l q^b \nonumber \\
\chi_3 &=& i \epsilon_{jkl} \bar{q}^a \gamma_4 \gamma_k B^{ab}_l q^b .
\label{eqn:chi2chi3defn}
\end{eqnarray}
These interpolating fields provide the strongest signal for the
$1^{-+}$ state.  In both fields the gluonic component is contributing
axial-vector quantum numbers.

\subsection{Lattice Field Strength Tensor}

In order to obtain the chromo-electric and chromo-magnetic fields with
which we build the hybrid operators, we make use of a modified version
of APE smearing, in which the smeared links do not involve averages
which include links in the temporal direction.  In this way we
preserve the notion of a Euclidean `time' and avoid overlap of the creation and
annihilation operators.  

In this study, the smearing fraction $\alpha = 0.7$ (keeping 0.3 of
the original link) and the process of smearing and $SU(3)$ link
projection is iterated six times \cite{Bonnet:2000dc}.
Each iteration of our modified APE-smearing algorithm proceeds as
\begin{eqnarray*}
\lefteqn{U_{i}(x) \to (1 - \alpha)\, U_i(x)} \nonumber \\
&+& \frac{\alpha}{4}\sum_{j=1}^{3} (1 - \delta_{ij}) \,U_j(x)\,U_i(x + \hat{j})\,U^\dag_j(x+ \hat{i}) \nonumber \\
&+& \frac{\alpha}{4}\sum_{j=1}^{3} (1 - \delta_{ij}) \,U^\dag_j(x -
\hat{j})\,U_i(x - \hat{j})\,U_j(x- \hat{j} + \hat{i}) \, ,\nonumber\\
\nonumber\\
\lefteqn{U_{4}(x) \to (1 - \alpha)\, U_4(x)} \nonumber \\
&+& \frac{\alpha}{6}\sum_{j=1}^{3} U_j(x)\,U_4(x + \hat{j})\,U^\dag_j(x+ \hat{4}) \nonumber \\
&+& \frac{\alpha}{6}\sum_{j=1}^{3} U^\dag_j(x - \hat{j})\,U_4(x -
\hat{j})\,U_j(x- \hat{j} + \hat{4}) \, . \nonumber\\
\end{eqnarray*}
Smearing the links permits the use of highly improved definitions of
the lattice field strength tensor, from which our hybrid operators are
derived.  Details of the ${\mathcal O(a^4)}$-improved tensor are given
in~\cite{Bilson-Thompson:2002jk}.  This amount of smearing is suitable
for the creation of exotic mesons.

%%%%%%%%%%%%%%%%%%%%%%%%%%%%%%%%%%%%%%%%%%%%%%%%%%%%%%%%%%%%%%%%%%%%%%%%%%
%
\begin{figure}[tbh]
\includegraphics[height=0.93\hsize,angle=90]{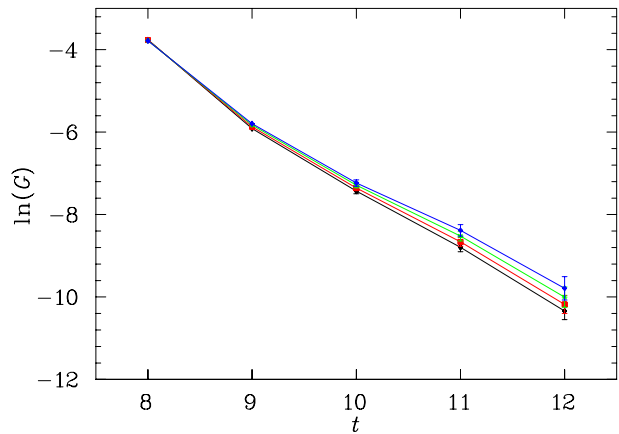}
\caption{\label{onemp_prop}Exotic meson propagator for interpolator $\chi_2$. Results are shown for every 2nd
quark mass in the simulation. Lower lines correspond to heavier quark masses. For all but the heaviest mass,
the signal is lost after t=12.}
\end{figure}

\begin{figure}[tbh]
\includegraphics[height=0.93\hsize,angle=90]{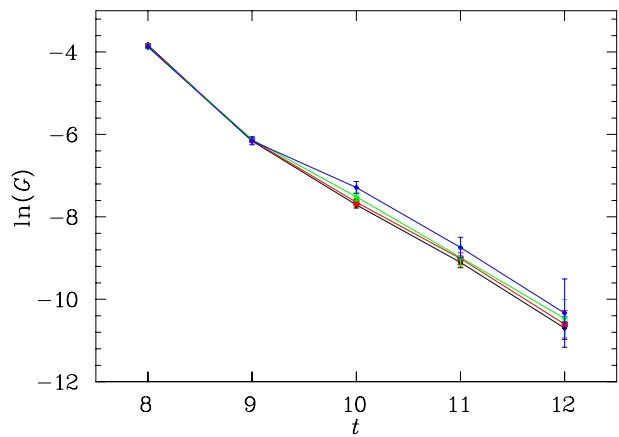}
\caption{\label{onemp_prop3}Exotic meson propagator for interpolator $\chi_3$. Results are shown for every 2nd
quark mass in the simulation. Lower lines correspond to heavier quark masses.}
\end{figure}

\subsection{Fat-Link Irrelevant Fermion Action}

Propagators are generated using the fat-link irrelevant clover (FLIC)
fermion action \cite{Zanotti:2001yb} where the irrelevant Wilson and
clover terms of the fermion action are constructed using fat links,
while the relevant operators use the untouched (thin) gauge links.
Fat links are created via APE smearing \cite{ape}.  In the FLIC
action, this reduces the problem of exceptional configurations
encountered with clover actions \cite{Boinepalli:2004fz}, and
minimizes the effect of renormalization on the action improvement
terms \cite{Leinweber:2002bw}.  Access to the light quark mass regime
is enabled by the improved chiral properties of the lattice fermion
action \cite{Boinepalli:2004fz}.  By smearing only the irrelevant,
higher dimensional terms in the action, and leaving the relevant
dimension-four operators untouched, short distance quark and gluon
interactions are retained.  Details of this approach may be found in
reference \cite{Zanotti:2001yb}.  FLIC fermions provide a new form of
nonperturbative ${\mathcal O}(a)$ improvement
\cite{Leinweber:2002bw,Boinepalli:2004fz} where near-continuum results
are obtained at finite lattice spacing.

%%%%%%%%%%%%%%%%%%%%%%%%%%%%%%%%%%%%%%%%%%%%%%%%%%%%%%%%%%%%%%%%%%%%%%%%%%
%
\begin{figure}[htb]
\includegraphics[height=0.93\hsize,angle=90]{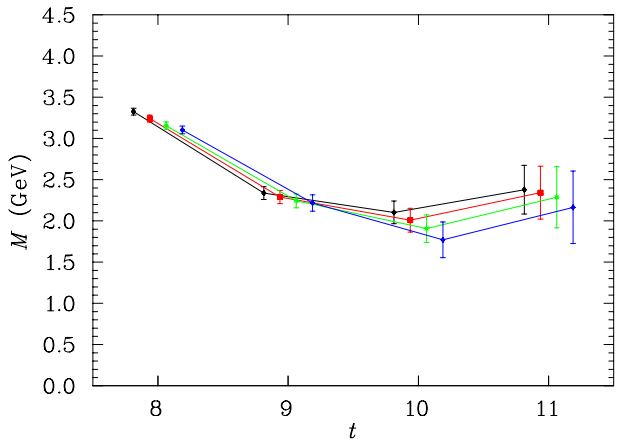}
\caption{\label{onemp}Effective mass for interpolator $\chi_2$. Plot symbols are as for the corresponding
propagator plot. }
\end{figure}

\begin{figure}[hbt]
\includegraphics[height=0.93\hsize,angle=90]{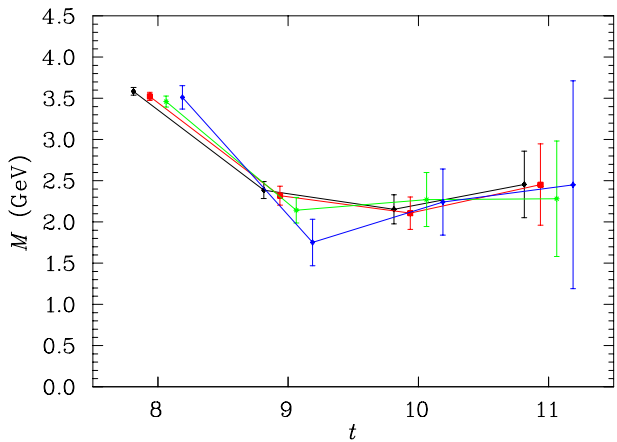}
\caption{\label{onempthree}As for Fig.~\ref{onemp}, but for interpolator $\chi_3$. Signal is lost after $t=11$.}
\end{figure}

\begin{figure}[t]
\includegraphics[height=0.93\hsize,angle=90]{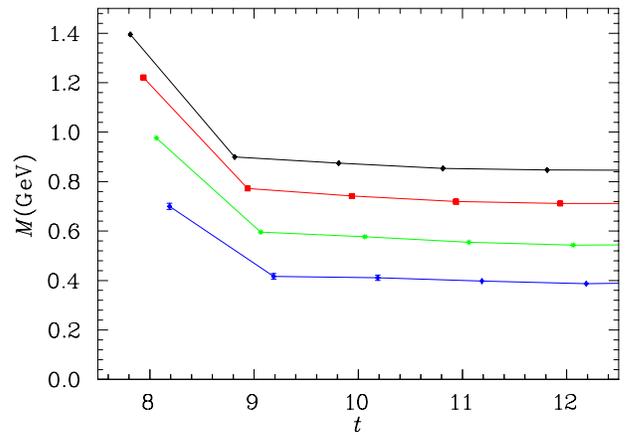}
\caption{\label{standardpion}Effective mass for the pseudoscalar pion interpolator $\bar{q} \gamma_5 q$.}
\end{figure}
\begin{figure}[t]
\includegraphics[height=0.93\hsize,angle=90]{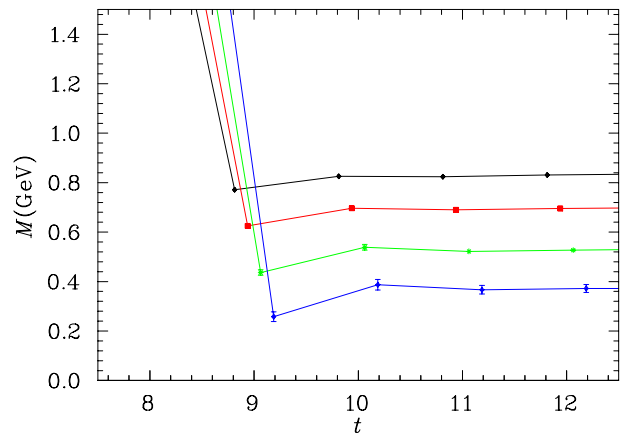}
\caption{\label{axialvectorpion}Effective mass for the axial-vector pion interpolator $\bar{q} \gamma_5 \gamma_4 q$.}
\end{figure}

\begin{figure}[htb]
\includegraphics[height=0.93\hsize,angle=90]{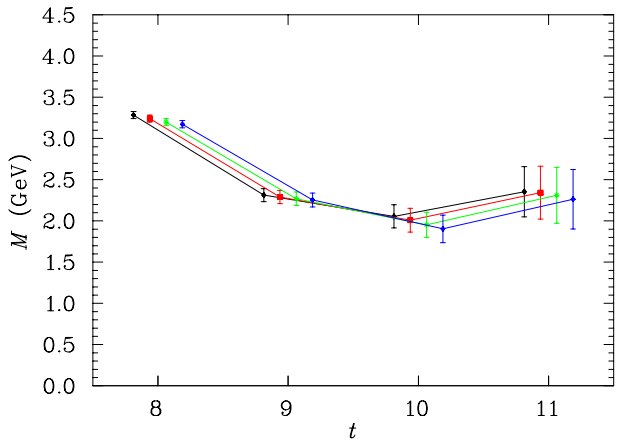}
\caption{\label{onemp_strange}Effective mass for the interpolator $\chi_2$ with a strange quark. }
\end{figure}

\begin{figure}[hbt]
\includegraphics[height=0.93\hsize,angle=90]{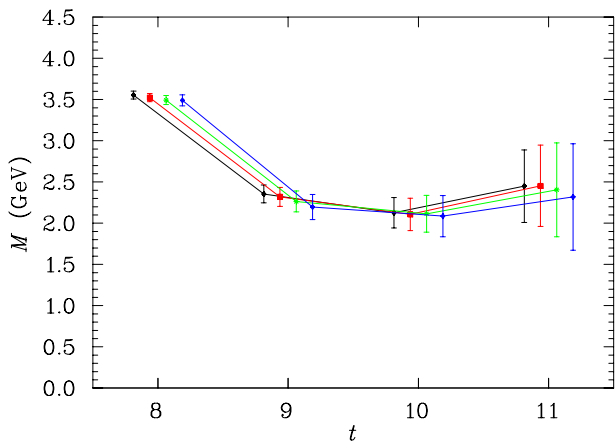}
\caption{\label{onempthree_strange}As for Fig.~\ref{onemp_strange}, but for interpolator $\chi_3$.}
\end{figure}

\begin{figure}[hbt]
\includegraphics[height=0.93\hsize,angle=90]{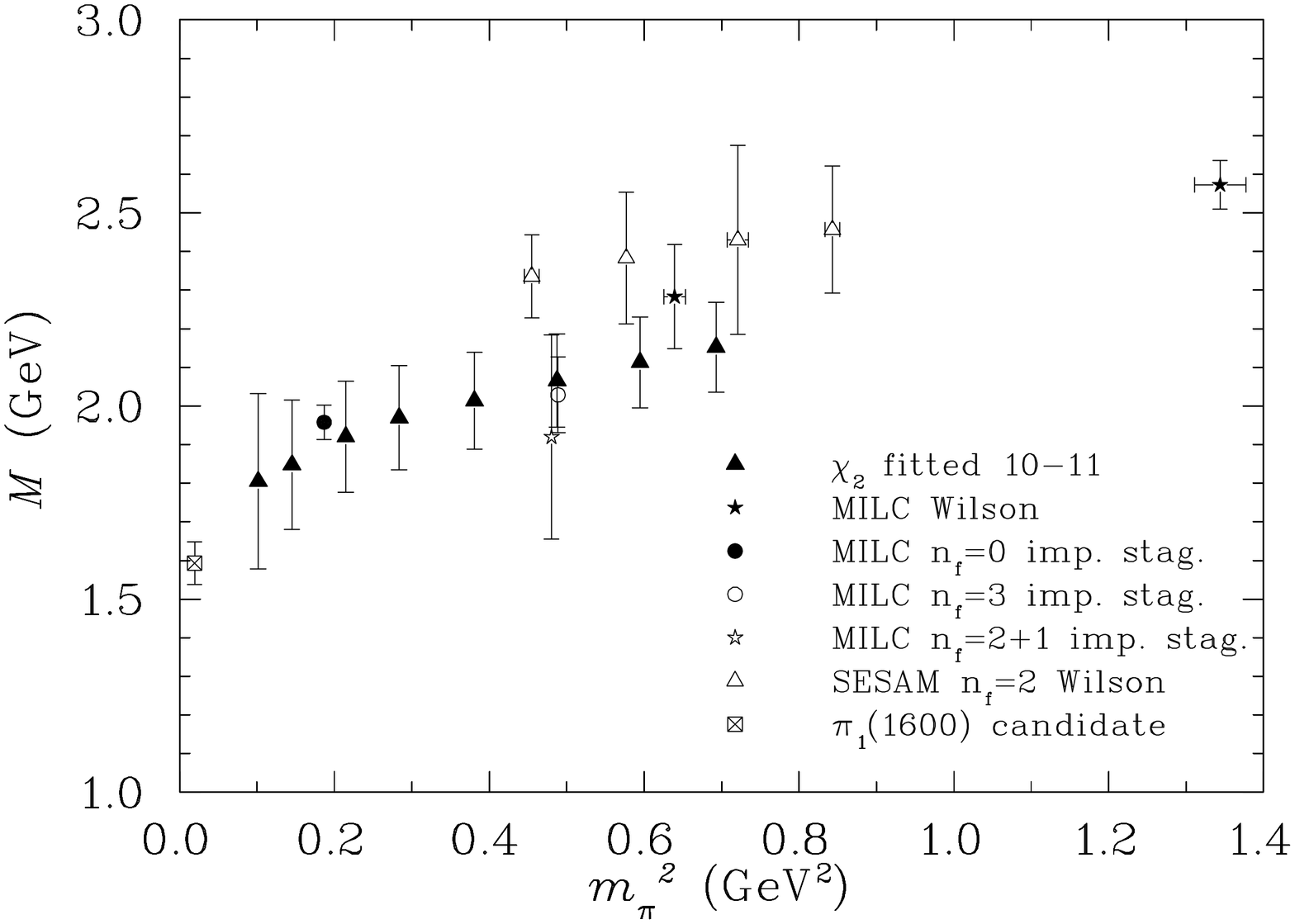}
\caption{\label{survey} A survey of results in this field. The MILC results are taken from ~\cite{Bernard:1997ib} and
show their $Q^4,1^{-+} \to 1^{-+}$ results, fitted from $t=3$ to $t=11$. Open and closed symbols denote dynamical and quenched simulations respectively. }
\end{figure}

\subsection{Gauge Action}

We use quenched-QCD gauge fields created by the CSSM Lattice
Collaboration with the ${\mathcal O}(a^2)$ mean-field improved
L\"uscher-Weisz plaquette plus rectangle gauge action
\cite{Luscher:1984xn} using the plaquette measure for the mean link.
The gauge-field parameters are defined by
\begin{eqnarray*}
S_G &=& \frac{5\beta}{3}\sum_{\scriptstyle x\, \mu\, \nu \atop
\scriptstyle \nu > \mu} 
\frac{1}{3}{\rm\ Re\ Tr} \left (1 - P_{\mu \nu}(x) \right) \\
&-&\frac{\beta}{12\, u_{0}^2}\sum_{\scriptstyle x\, \mu\, \nu\, \atop
\scriptstyle \nu > \mu}
\frac{1}{3}{\rm\ Re\ Tr} \left (2 - R_{\mu \nu}(x) \right ) \, ,
\label{gaugeaction}
\end{eqnarray*}
where $P_{\mu \nu}$ and $R_{\mu \nu}$ are defined in the usual manner
and the link product $R_{\mu \nu}$ contains the sum of the
rectangular $1\times 2$ and $2 \times 1$ Wilson loops.  

The CSSM configurations are generated using the Cabibbo-Marinari
pseudo-heat-bath algorithm~\cite{Cabibbo:1982zn} using a parallel algorithm
with appropriate link partitioning \cite{Bonnet:2000db}.  To improve
the ergodicity of the Markov chain process, the three diagonal SU(2)
subgroups of SU(3) are looped over twice~\cite{Bonnet:2001rc} and a
parity transformation \cite{Leinweber:2003sj} is applied randomly to
each gauge field configuration saved during the Markov chain process.

\subsection{Simulation Parameters}

The calculations of meson masses are performed on $20^3\times 40$
lattices at $\beta=4.53$, which provides a lattice spacing of $a =
0.128(2)$~fm set by the Sommer parameter $r_0=0.49\ \rm fm$.

A fixed boundary condition in the time direction is used for the fermions
by setting $U_t(\vec x, N_t) = 0\ \forall\ \vec x$ in the hopping terms
of the fermion action, with periodic boundary conditions imposed in the
spatial directions.

Eight quark masses are considered in the calculations and the strange
quark mass is taken to be the third heaviest quark mass.  This
provides a pseudoscalar mass of 697 MeV which compares well with the
experimental value of $( 2M_K^2 - M_\pi^2 )^{1/2} = 693\, {\rm MeV}$
motivated by leading order chiral perturbation theory.  

The analysis is based on a sample of 345 configurations, and the error
analysis is performed by a third-order single-elimination jackknife,
with the $\chi^2$ per degree of freedom (${\chi^2}/dof$) obtained via
covariance matrix fits.

\section{Results}

%%%%%%%%%%%%%%%%%%%%%%%%%%%%%%%%%%%%%%%%%%%%%%%%%%%%%%%%%%%%%%%%%%%%%%%%%%
%
\begin{table*}[hbt]
\caption{\label{masstable}$1^{-+}$ Exotic Meson mass $m$ (GeV) vs square
  of pion mass $m_\pi^2$ (GeV$^2$).} 
\begin{ruledtabular}
\begin{tabular}{clllllll}
$m_{\pi}^2$ & \multicolumn{2}{c}{$\chi_2$ fit 10-11} & \multicolumn{2}{c}{$\chi_2$ fit 10-12} & \multicolumn{2}{c}{$\chi_3$ fit 10-11} \\
& $m$ & $\chi^2/dof$ & $m$ & $\chi^2/dof$ & $m$ & $\chi^2/dof$ \\
\hline
0.693(3) & 2.15(12) & 0.69 & 2.16(11) & 0.44 & 2.20(15) & 0.45 \\
0.595(4) & 2.11(12) & 0.77 & 2.12(11) & 0.51 & 2.18(16) & 0.46\\
0.488(3) & 2.07(12) & 0.85 & 2.08(12) & 0.59 & 2.15(17) & 0.41\\
0.381(3) & 2.01(12) & 0.91 & 2.03(12) & 0.65 & 2.14(19) & 0.29\\
0.284(3) & 1.97(13) & 0.78 & 1.98(13) & 0.55 & 2.27(29) & 0.00012\\
0.215(3) & 1.92(14) & 0.78 & 1.92(14) & 0.40 & 2.25(31) & 0.02\\
0.145(3) & 1.85(17) & 0.57 & 1.84(17) & 1.76 & 2.26(37) & 0.02\\
0.102(4) & 1.80(23) & 0.13 & 1.75(23) & 3.04 & 2.46(58) & 0.03\\
\end{tabular}
\end{ruledtabular}
\end{table*}

\begin{table*}[hbt]
\caption{\label{masstable_strange}Strangeness $\pm 1\, \, 1^{-}$ Meson mass $m$ (GeV) vs square of pion mass $m_\pi^2$(GeV$^2$).}
\begin{ruledtabular}
\begin{tabular}{clllllll}
$m_{\pi}^2$ & \multicolumn{2}{c}{$\chi_2$ fit 10-11} & \multicolumn{2}{c}{$\chi_2$ fit 10-12} & \multicolumn{2}{c}{$\chi_3$ fit 10-11} \\
& $m$ & $\chi^2/dof$ & $m$ & $\chi^2/dof$ & $m$ & $\chi^2/dof$ \\
\hline
0.693(3) & 2.11(12) & 0.76& 2.12(11) & 0.51 & 2.17(16) & 0.44\\
0.595(4) & 2.09(12) & 0.81 & 2.10(12) & 0.55 & 2.16(16) & 0.44\\
0.488(3) & 2.07(12) & 0.85 & 2.08(12) & 0.59 & 2.15(17) & 0.41\\
0.381(3) & 2.04(12) & 0.88 & 2.05(12) & 0.63 & 2.15(18) & 0.36\\
0.284(3) & 2.01(13) & 0.85 & 2.02(12) & 0.63 & 2.25(20) & 0.22\\
0.215(3) & 1.99(13) & 0.87 & 2.00(12) & 0.64 & 2.11(20) & 0.29\\
0.145(3) & 1.97(13) & 0.73 & 1.97(13) & 0.54 & 2.12(22) & 0.11\\
0.102(4) & 1.96(14) & 0.56 & 1.96(14) & 0.39 & 2.09(24) & 0.01\\
\end{tabular}
\end{ruledtabular}
\end{table*}

Figures \ref{onemp_prop} and \ref{onemp_prop3} show the natural log of
the correlation functions calculated with interpolators $\chi_2$ and
$\chi_3$ from Eq.~\ref{eqn:chi2chi3defn} respectively.  The curves become linear after two time slices
from the source, corresponding to approximately $0.256$~fm. This is
consistent with Ref.~\cite{Bernard:1997ib}, where a similar effect is
seen after approximately 3 to 4 time slices, corresponding to $0.21$ to
$0.28$~fm following the source.

Figures~\ref{onemp} and~\ref{onempthree} show the effective mass for
the two different interpolators.  For clarity, we have plotted the
results for every second quark mass used in our simulation.  The
plateaus demonstrate that we do indeed see an exotic signal in
quenched lattice QCD.  This is significant, as we expect the two
interpolating fields to possess considerably different excited-state
contributions, based on experience with pseudoscalar interpolators
\cite{Holl:2005vu}.

For example, the approach to the pion mass plateau is from above
(below) for the pseudo-scalar (axial-vector) interpolating field as
illustrated in Fig.~\ref{standardpion} (Fig.~\ref{axialvectorpion}).
This exhibits the very different overlap of the interpolators with
excited states.  As in the $1^{-+}$ interpolators, the role of
$\gamma_4$ in the pion interpolators is to change the sign with which
the large-large and small-small spinor components are combined.  
We also present results for the strangeness $\pm 1$ analogue of the
$1^{-+}$ in Figs.~\ref{onemp_strange} and \ref{onempthree_strange}.

Table~\ref{masstable} summarizes our results for the mass of the
$1^{-+}$ meson, with the squared pion-mass provided as a measure of
the input quark mass.  The agreement observed in the results obtained
from the two different $1^{-+}$ hybrid interpolators provides evidence
that a genuine ground-state signal for the exotic has been observed.

Table~\ref{masstable_strange} summarizes our results for the mass
of the strangeness $\pm\,1$, $J^{P} = 1^{-}$ meson.

Finally, in Fig.~\ref{survey} we summarize a collection of results for
the mass of $1^{-+}$ obtained in lattice QCD simulations thus far.
The current results presented herein are compared with results from
the MILC \cite{Bernard:1997ib,Bernard:2002rz} and SESAM \cite{Lacock:1998be}
collaborations, both of which provide a consistent scale via $r_0$.

Our results compare favorably with earlier work at large quark masses.
Agreement within one sigma is observed for all the quenched simulation
results illustrated by filled symbols.  It is interesting that the
dynamical Wilson fermion results of the SESAM collaboration
\cite{Lacock:1998be} tend to sit somewhat higher as this is a well
known effect in baryon spectroscopy \cite{Young:2002cj,Zanotti:2001yb,Zanotti:2002ax,Zanotti:2004dr}.  

\section{Physical Predictions}

In comparing the results of quenched QCD simulations with experiment,
the most common practice is to simply extrapolate the results linearly
in $m_q$ or $m_\pi^2$ to the physical values.  However, such an
approach provides no opportunity to account for the incorrect chiral
nonanalytic behavior of quenched QCD
\cite{Young:2002cj,Young:2004tb,Leinweber:2004tc,Leinweber:2005jb}.

Unfortunately, little is known about the chiral nonanalytic behavior
of the $1^{-+}$ meson.  Ref.~\cite{Thomas:2001gu} provides a full QCD
exploration of the chiral curvature to be expected from transitions to
nearby virtual states and channels which are open at physical quark
masses.  While virtual channels act to push the lower-lying
single-particle $1^{-+}$ state down in mass, it is possible to have
sufficient strength lying below the $1^{-+}$ in the decay channels
such that the $1^{-+}$ mass is increased
\cite{Leinweber:1993yw,Allton:2005fb}.  Depending on the parameters
considered in Ref.~\cite{Thomas:2001gu} governing the couplings of the
various channels, corrections due to chiral curvature are estimated at
the order of $+20$ to $-40$ MeV.

Generally speaking, chiral curvature is suppressed in the quenched
approximation.  For mesons, most of the physically relevant diagrams
involve a sea-quark loop and are therefore absent
\cite{Allton:2005fb,Sharpe:1992ft}.  However, the light quenched
$\eta'$ meson can provide new nonanalytic behavior, with the lowest
order contributions coming as a negative-metric contribution through
the double-hairpin diagrams.  Not only do these contributions alter
the $1^{-+}$ mass through self-energy contributions, but at
sufficiently light quarks masses, open decay channels can dominate the
two-point correlator and render its sign negative.

For the quenched $1^{-+}$ meson, the $a_1 \eta^{\prime}$ channel can
be open.  Using the pion mass as the $\eta'$ mass a direct calculation
of the mass of an $a_1 \eta^{\prime}$ two-particle state indicates
that the $1^{-+}$ hybrid lies lower than the two-particle state for
heavy input quark mass.  This indicates that the hybrid interpolator
is effective at isolating a single-particle bound state as opposed to the
two-particle state at heavy quark masses.  This is particularly true for the 
case here, where long Euclidean time evolution is difficult. 

\begin{figure}[t]
\includegraphics[height=0.93\hsize,angle=90]{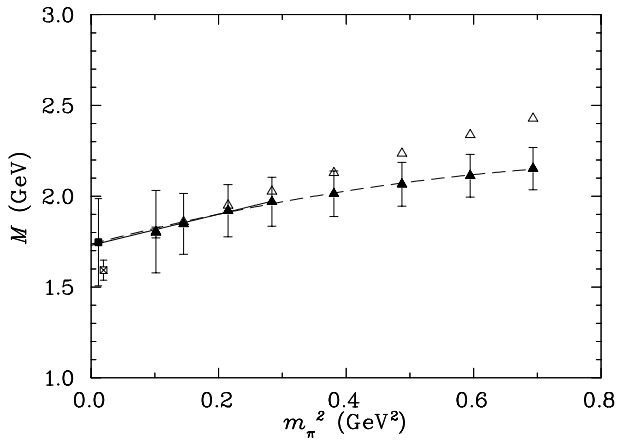}
\caption{\label{extrapolation}The $1^{-+}$ exotic meson mass obtained
  from fits of the effective mass of the hybrid interpolator $\chi_2$
  from $t=10 \to 12$ (full triangles) are compared with the $a_1
  \eta^{\prime}$ two-particle state (open triangles).  The
  extrapolation curves include a quadratic fit to all eight quark
  masses (dashed line) and a linear fit through the four lightest
  quark masses (solid line).  The full square is result of linear
  extrapolation to the physical pion mass, while the open square
  (offset for clarity) indicates the $\pi_1(1600)$ experimental
  candidate. .}
\end{figure}

As the light quark mass regime is approached, the trend of the one and
two-particle states illustrated in Fig.~\ref{extrapolation}, suggests
that they either merge or cross at our second lightest quark mass, such that
the exotic $1^{-+}$ may be a resonance at our lightest quark mass and at the
physical quark masses.  We note that the exotic $1^{-+}$ mass displays
the common resonance behavior of becoming bound at quark masses
somewhat larger than the physical quark masses. This must happen at sufficiently
heavy quark masses by quark counting rules, i.e $2q \to 4q$ for the $1^{-+}$ to
$a_1 \eta^{\prime}$ transition. 

One might have some concerns about $a_1 \eta^{\prime}$
contaminations in the two-point correlation function affecting the
extraction of the $1^{-+}$ meson mass \cite{Liu:2005la}. However we can
already make some comments.  

Under the assumption that the coupling to the quenched $a_1 \eta^{\prime}$ channel
comes with a negative metric, as suggested by chiral perturbation theory arguments,
and from the observation that our correlation functions are positive, then it would
appear that our interpolators couple weakly to the decay channel.
Furthermore, at heavy quark masses the correlation function is dominated by the
$1^{-+}$ bound state already at early Euclidean times suggesting that coupling to the
decay channel is weak.

Thus we conclude that the hybrid interpolating fields used to explore
the $1^{-+}$ quantum numbers are well-suited to isolating the
single-particle $1^{-+}$ exotic meson.

Moreover, since the mass of the $a_1 \eta^{\prime}$ channel is similar
or greater than the single-particle $1^{-+}$ state, one can conclude
that the double-hairpin $a_1 \eta^{\prime}$ contribution to the self
energy of the single-particle $1^{-+}$ exotic meson is repulsive in
quenched QCD.  Since the curvature observed in Fig.~\ref{survey}
reflects attractive interactions, we can also conclude that quenched
chiral artifacts are unlikely to be large.

\begin{figure}[t]
\includegraphics[height=0.93\hsize,angle=90]{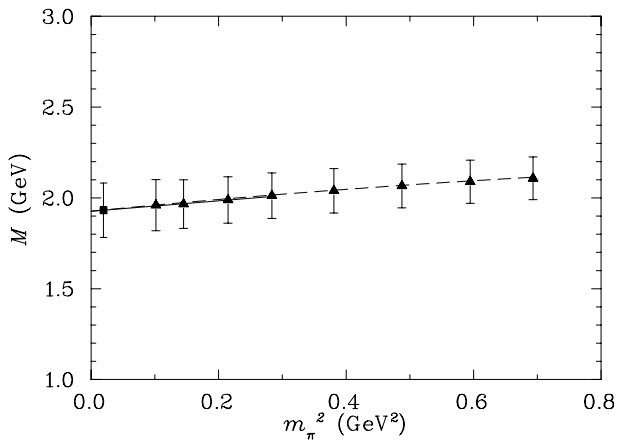}
\caption{\label{extrapolation_strange} Extrapolation of the associated
  strangeness $\pm 1$ $J^P = 1^{-}$ state obtained from $\chi_2$.
  Symbols are as in Fig.~\ref{extrapolation}.}
\end{figure}

Hence we proceed with simple linear and quadratic extrapolations in
quark mass to the physical pion mass, with the caution that chiral
nonanalytic behavior could provide corrections to our simple
extrapolations the order of 50 MeV in the $1^{-+}$ mass\cite{Thomas:2001gu}.

Figures \ref{extrapolation} and \ref{extrapolation_strange} illustrate
the extrapolation of the $1^{-+}$ exotic and its associated
strangeness $\pm1$ $1^{-}$ state to the limit of physical quark mass.
We perform the linear fit using the four lightest quark masses and
fit the quadratic form to all 8 masses.  A third-order
single-elimination jackknife error analysis yields masses of $1.74(24)$
and $1.74(25)$ GeV for the linear and quadratic fits, respectively.
These results agree within one standard deviation with the
experimental $\pi_1 (1600)$ result of $1.596^{+25}_{-14}\ {\rm GeV}$, and exclude the
mass of the $\pi_1 (1400)$ candidate.

The associated parameters of the fits are as follows.  The linear
form
\begin{eqnarray*}
m_{1^{-+}} &=& a_0 + a_2\, m_{\pi}^2 \, ,
\end{eqnarray*}
yields best fit parameters of 
\begin{eqnarray*}
&a_0& = 1.73 \pm {0.15}\ {\rm GeV}\, , \nonumber\\
&a_2& = 0.85 \pm {0.35}\ {\rm GeV}^{-1} \, .
\end{eqnarray*}
The quadratic fit, with formula
\begin{eqnarray*}
m_{1^{-+}} &=& a_0 + a_2\, m_{\pi}^2 + a_4\, m_{\pi}^4 \, ,
\end{eqnarray*}
returns parameters 
\begin{eqnarray*}
&a_0& = +1.74 \pm{0.15}\ {\rm GeV}, \nonumber \\
&a_2& = +0.91 \pm{0.39}\ {\rm GeV}^{-1}, \nonumber \\
&a_4& = -0.46 \pm{0.35}\ {\rm GeV}^{-3} \, .
\end{eqnarray*}

%%%%%%%%%%%%%%%%%%%%%%%%%%%%%%%%%%%%%%%%%%%%%%%%%%%%%%%%%%%%%%%%%%%%%%%%%%
\section{Conclusion}

We have found a compelling signal for the $J^{PC}=1^{-+}$ exotic meson
, from which we can extrapolate a physical
mass of $1.74(24)$ GeV.  Thus for the first time in lattice studies,
we find a $1^{-+}$ mass in agreement with the $\pi_1 (1600)$
candidate.

The $\chi_2$ interpolating field appears to be extremely useful for avoiding
contamination from the $a_1 \eta^{\prime}$ channel, and thus is an
excellent choice for this kind of study.

We have also presented the first results for a strangeness $\pm 1$
partner of the exotic $1^{-+}$ meson lying at $1.92(15)$ GeV.  

Looking forward, it will be important to quantify the effects of the
quenched approximation.  We plan to revisit these calculations at some
future point using full dynamical FLIC fermions
\cite{Kamleh:2004xk,Kamleh:2003wb}.  Of particular interest will be
the extent to which the curvature observed in approaching the chiral
regime is preserved in full QCD.

Additionally, we intend to explore the dependence of the exotic signal
on the nature of the fermion source.  Whilst the rapidity with which
we establish a plateau in our effective mass plots suggests that our
current fermion operator smearing is near optimal for isolating the
ground state, it might be possible to reduce the statistical errors
through a careful selection of parameters coming out of a systematic
exploration of the parameter space.

Finally, a detailed finite volume analysis should be performed in order
to further explore the role of the two-body decay channel.

%%%%%%%%%%%%%%%%%%%%%%%%%%%%%%%%%%%%%%%%%%%%%%%%%%%%%%%%%%%%%%%%%%%%%%%%%%
\begin{acknowledgments}

We thank Doug Toussaint for sharing his collection of results for the
$1^{-+}$.  We thank the Australian Partnership for Advanced Computing
(APAC) and the Australian National Computing Facility for Lattice
Gauge Theory managed by the South Australian Partnership for Advanced
Computing (SAPAC) for generous grants of supercomputer time which have
enabled this project.  This work was supported by the Australian
Research Council.

\end{acknowledgments}

\end{document}